\documentclass[reqno,12pt]{amsart}
\usepackage{fullpage}
\usepackage{amsmath,amsfonts,amssymb}

\numberwithin{equation}{section}

\newtheorem{theorem}{Theorem}[section]{\bf}{\it}

\newtheorem{lemma}[theorem]{Lemma}{\bf}{\it}

{\bf}{\it}

\newtheorem{proposition}[theorem]{Proposition}{\bf}{\it}

\def\Ref#1{Ref.~\cite{#1}}

\def\eos/{equation of state}
\def\esos/{equations of state}
\def\com/{constant of motion}
\def\csom/{constants of motion}

\def\Rnum{{\mathbb R}}

\def\t{\mathrm{t}}

\def\Esp{{\mathcal E}}
\def\Hop{{\mathcal H}}
\def\Rop{{\mathcal R}}
\def\Dop{{\mathcal D}}

\def\X{{\mathrm X}}
\def\pr{\mathrm{pr}}

\def\I{{\mathcal I}}

\def\F{{F}}
\def\G{{G}}
\def\k{\kappa}
\def\A#1{\mathrm{A}_{#1}}

\def\gas{\text{gas}}

\def\ii{5} 
\def\iii{3} 
\def\iv{7} 
\def\v{9} 
\def\vi{4} 
\def\vii{8} 
\def\viii{6} 
\def\ix{10} 
\def\vvi{{11}} 

\allowdisplaybreaks[3]
\begin{document}

\title{Symmetries and Casimirs of radial\\ compressible fluid flow and gas dynamics\\ in $n>1$ dimensions}

\author{
Stephen C. Anco
\lowercase{\scshape{and}} 
Sara Seifi
\lowercase{\scshape{and}} 
Thomas Wolf
\\\lowercase{\scshape{
Department of Mathematics and Statistics\\
Brock University\\
St. Catharines, ON Canada}}
}

\begin{abstract}
Symmetries and Casimirs are studied 
for the Hamiltonian equations of radial compressible fluid flow in $n>1$ dimensions. 
An explicit determination of all Lie point symmetries is carried out, 
from which a complete classification of all maximal Lie symmetry algebras is obtained. 
The classification includes all Lie point symmetries that exist only for special \esos/. 
For a general \eos/, 
the hierarchy of advected conserved integrals found in recent work 
is proved to consist of Hamiltonian Casimirs. 
A second hierarchy that holds only for an entropic \eos/ is explicitly shown to comprise 
non-Casimirs which yield a corresponding hierarchy of generalized symmetries
through the Hamiltonian structure of the equations of radial fluid flow. 
The first-order symmetries are shown to generate a non-abelian Lie algebra. 
Two new kinematic conserved integrals found in recent work 
are likewise shown to yield additional first-order generalized symmetries
holding for a barotropic \eos/ and an entropic \eos/. 
These symmetries produce an explicit transformation group acting on solutions of 
the fluid equations. 
Since these equations are well known 
to be equivalent to the equations of gas dynamics,
all of the results obtained for $n$-dimensional radial fluid flow 
carry over to radial gas dynamics. 
\end{abstract}

\maketitle

\section{Introduction}

In recent work \cite{AncSeiDar} 
studying radial compressible fluid flow in $n>1$ dimensions, 
new conserved integrals and new advected scalars (invariants) have been found 
which are not inherited by radial reduction from the known 
conserved integrals and invariants of $n$-dimensional non-radial fluid flow
\cite{AncDar2009,AncDar2010}. 
These ``hidden'' quantities indicate that, 
compared to the full Euler equations governing compressible fluid flow,
the radial Euler equations have a much richer structure. 
The same results hold for the equations of gas dynamics, 
which have a well-known equivalence \cite{Whi-book} to the Euler equations of compressible fluid flow. 

Two of the new conserved integrals describe kinematic quantities,
one being an enthalpy-flux which holds for barotropic \esos/,
and the other being an entropy-weighted energy which holds for entropic \esos/. 
Most interestingly, 
the new advected scalars comprise two infinite hierarchies that hold respectively 
for general non-barotropic \esos/ and entropic \esos/. 
Both hierarchies are generated by a recursion operator 
applied to basic advected scalars. 
Each of the hierarchies gives rise to corresponding advected integrals on transported radial domains, 
which are obtained in terms of conserved densities derived from the advected scalars. 

These unexpected results motivate the present study of 
the Hamiltonian structure, Casimirs, and symmetries 
of the equations for radial fluid flow and radial gas dynamics in $n>1$ dimensions,
with a general \eos/. 

Firstly,
all point symmetries will be determined, 
including any that exist only for special \esos/. 
A rich structure of symmetry algebras and attendant \esos/ is seen to exist,
which is far wider than the structure of kinematic conserved integrals
found in \Ref{AncSeiDar}. 
The corresponding symmetry transformation groups will be described. 

Secondly, 
all Hamiltonian Casimirs up to first-order will be determined. 
It is found that these coincide with the new advected integrals of zeroth and first order 
in the hierarchy holding for a general \eos/. 
An inductive proof that all of the higher-order advected integrals are Casimirs is given. 

Thirdly, 
the remaining new advected integrals are shown to not be Casimirs. 
They instead give rise to generalized (non-point) symmetries 
through the general well-known correspondence 
between non-Casimir conserved integrals and Hamiltonian symmetries. 
The resulting symmetries of the radial compressible fluid equations 
include subalgebras of first-order symmetries and higher-order symmetries. 
This is a ``hidden'' symmetry structure which does not arise from radial reduction of 
the Hamiltonian symmetries admitted by the $n$-dimensional Euler equations for compressible fluid flow. 
The first-order symmetries are shown to produce an explicit transformation group
acting on solutions of the radial equations. 

The main results presented here can have 
physical applications to explosive flows and implosive flows
(see e.g. \cite{Tay,Kel,Wel,JenTsi,Whi-book,Can-book}), 
as well as numerous engineering applications
(see e.g. \cite{Aun-book,CasRob-book}). 
Moreover, they may be indicative of hidden structure for more general zero-vorticity flows. 
 
The rest of this paper is organized as follows. 
In Section~\ref{sec:eqns}, 
the equations of radial fluid flow and their equivalence to the equations of radial gas dynamics are summarized. 
The Hamiltonian structure of these equations and attendant properties are presented. 
In Section~\ref{sec:pointsymms}, 
the classification of point symmetries and algebras is stated.
The results on Hamiltonian Casimirs are presented in Section~\ref{sec:casimirs}.
In Section~\ref{sec:Hamilsymms},
the Hamiltonian symmetries are derived. 
Finally, some concluding remarks are made in Section~\ref{sec:remarks}. 

Appendix~\ref{app:computation} provides remarks on computational aspects of the main classifications. 
Appendix~\ref{app:identities} summarizes some variational identities 
which are used in the proofs, 
as well as some basic definitions. 

General results on symmetries can be found in \Ref{Olv-book,BCA,Anc-review}. 
See \Ref{Arn1969,Dez,Ser,Ver,KheChe,Kur,Kup-book,ArnKhe-book} for key work on the 
Hamiltonian structure and Casimirs for the non-radial Euler equations. 
The terminology and notation in the present work is the same as in \Ref{AncSeiDar}, 
which comes from \Ref{Olv-book} for the material on 
symmetries and Hamiltonian structures, 
and from \Ref{AncDar2009,AncDar2010,AncWeb2020} for the material on 
conserved integrals and advected quantities. 
The mathematically setting will be calculus in jet space \cite{Olv-book,Anc-review}.

\section{Radial flow equations and Hamiltonian structure}\label{sec:eqns}

The radial reduction of the Euler equations of compressible fluid flow in $n$ dimensions without boundaries 
is given by the system 
\begin{align}
& 
U_t+ U U_r + ( p_S S_r + p_{\rho} \rho_r )/\rho =0, 
\label{U.eqn}
\\
& 
\rho_t + (U\rho)_r +\tfrac{n-1}{r} U \rho = 0, 
\label{rho.eqn}
\\
& 
S_t + U S_r = 0, 
\label{S.eqn}
\end{align}
where $U=U(t,r)$ is the radial component of the fluid velocity, 
$\rho=\rho(t,r)$ is the fluid density, 
and $S=S(t,r)$ is the local entropy. 

To close this system,  an \eos/ needs to be specified, which in general is given by 
\begin{equation}
p=p(\rho,S) .
\label{eos}
\end{equation}
The space of solutions $(U(t,r),\rho(t,r),S(t,r))$ of the closed system \eqref{U.eqn}, \eqref{rho.eqn}, \eqref{S.eqn}, \eqref{eos} will be denoted as $\Esp$. 

The \eos/ \eqref{eos} determines all thermodynamic quantities 
in terms of the internal energy $e(\rho,S)$ (per unit mass) 
through the thermodynamic relation 
$T\,dS =de + p\,d(1/\rho)$, where $T$ is the local temperature (per unit mass). 
In particular, 
the internal energy is given by 
\begin{equation}\label{e}
e(\rho,S) =\int (p(\rho,S)/\rho^2)\,d\rho , 
\end{equation}
which determines 
\begin{equation}
T(\rho,S) =\frac{\partial e}{\partial S}\Big|_{\rho} = \int (p_S(\rho,S)/\rho^2)\,d\rho . 
\end{equation}

The equations governing radial gas dynamics consist of 
the density equation \eqref{rho.eqn}
and the velocity equation formulated as 
\begin{equation}\label{U.gaseqn}
U_t + U U_r + p_r/\rho =0 , 
\end{equation}
together with the pressure equation 
\begin{equation}\label{p.gaseqn}
p_t + U p_r + a^2 \rho (U_r +\tfrac{n-1}{r} U) = 0 , 
\end{equation}
where 
\begin{equation}\label{a.gaseqn}
a=a(\rho,p) >0
\end{equation}
is the speed of sound. 
The pressure equation can be derived from the \eos/ \eqref{eos}
by use of the implicit function theorem to obtain $S=F(\rho,p)$, 
which is then substituted into the entropy equation \eqref{S.eqn}
and simplified using the density equation \eqref{rho.eqn}, 
with
\begin{equation}
a^2=-F_\rho/F_p = \frac{\partial p}{\partial\rho}\Big|_{S=F(\rho,p)} . 
\label{asq.rel}
\end{equation}
Conversely, the entropy equation can be recovered from the pressure equation \eqref{p.gaseqn} 
by solving $F_\rho + a^2(\rho,p) F_p=0$ to obtain $S=F(\rho,p)$, 
which is then observed to satisfy the entropy equation 
via the density and pressure equations. 

As a consequence,
the \esos/ familiar in fluid flow --- barotropic, polytropic, ideal gas --- 
and in gas dynamics --- ideal gas law --- 
have a direct correspondence 
in the equivalent formulations \eqref{U.eqn}--\eqref{eos} and \eqref{rho.eqn},\eqref{U.gaseqn}--\eqref{a.gaseqn}. 
Explicit equivalences are given in \Ref{AncSeiDar}.

\subsection{Radial Hamiltonian formulation}

The radial Euler equations \eqref{U.eqn}--\eqref{S.eqn} 
possess a Hamiltonian formulation which arises directly from 
reduction of the well-known Hamiltonian formulation \cite{Ver,AncDar2010} of 
the $n$-dimensional Euler equations. 
It is given by 
\begin{equation}
\partial_t 
\begin{pmatrix}
U \\ \rho \\ S 
\end{pmatrix}
= \Hop 
\begin{pmatrix}
\delta H/\delta U \\ \delta H/\delta \rho \\ \delta H/\delta S 
\end{pmatrix}
\label{Hamil.eqns}
\end{equation}
with the Hamiltonian 
\begin{equation}
H = \int_0^\infty \rho (\tfrac{1}{2} U^2 +e) r^{n-1}\,dr
\label{Hamil}
\end{equation}
where 
\begin{equation}
\Hop =
\begin{pmatrix}
0 &  -D_r r^{1-n} &  r^{1-n}\frac{1}{\rho} S_r \\
-r^{1-n} D_r & 0 & 0 \\
-r^{1-n}\frac{1}{\rho} S_r & 0 & 0 
\end{pmatrix}
\label{Hamil.op}
\end{equation}
is the (non-canonical) Hamiltonian operator, also known as a co-symplectic operator. 
Here $D_r$ denotes the total $r$-derivative (see Appendix~\ref{app:identities}). 
A general discussion of the properties of Hamiltonian (co-symplectic) operators 
can be found in \Ref{Olv-book}. 

The Hamiltonian \eqref{Hamil} physically describes a conserved energy
for radial fluid flow. 

This Hamiltonian structure \eqref{Hamil.eqns}--\eqref{Hamil.op}
corresponds to a Poisson bracket. 
Specifically, 
for any two functionals $\F$ and $\G$ on the whole radial domain $(0,\infty)$, 
their Poisson bracket is defined in terms of Hamiltonian operator \eqref{Hamil.op} by 
\begin{equation}
\{\F,\G\} = \int_0^\infty\big( \nabla_{U,\rho,S}^\t \F\Hop \nabla_{U,\rho,S} \G \big) r^{n-1}\, dr
\label{PB}
\end{equation}
modulo a trivial functional, 
where 
$\nabla_{U,\rho,S}   = \begin{pmatrix}
\frac{\delta}{\delta U} ,\ \frac{\delta}{\delta \rho},\ \frac{\delta}{\delta S}
\end{pmatrix}$
denotes the variational gradient,
and the ``$\t$'' denotes the transpose. 
Note that a functional is trivial if its density is a total $r$-derivative 
whereby the functional reduces identically to boundary terms at $r=0$ and $r\to\infty$
which will vanish under suitable conditions. 

The Poisson bracket \eqref{PB} is a bilinear map 
and has the properties that it is skew and obeys the Jacobi identity,
which follow from general results about Hamiltonian operators \cite{Olv-book}. 

In principle, the use of the radial domain $(0,\infty)$ could be replaced by 
a transported radial domain $V(t)$, whose points $r(t)$ satisfy 
$\frac{d}{dt}r(t) = U(r(t),t)$. 
The resulting bracket defined by 
$\{\F,\G\} = \int_{V(t)}\big( \nabla_{U,\rho,S}^\t \F\Hop \nabla_{U,\rho,S} \G \big) r^{n-1}\, dr$
will be skew and continue to obey the Jacobi identity, 
since the variation of $V(t)$ will contribute only boundary terms. 
More development of this approach will be pursued elsewhere. 

If a non-trivial functional $\F$ on the whole radial domain 
has no explicit dependence on $t$, 
then its time evolution is given by 
\begin{equation}
\dfrac{d\F}{dt} = \{F,H\} 
\label{time.evol}
\end{equation}
modulo a trivial functional. 
Hence, such a non-trivial functional $\F$ will be a conserved integral 
if and only if $\{\F,H\}=0$. 

Radial gas dynamics has a similar Hamiltonian formulation, 
which can be derived through 
the change of variables $(\rho,S)\to (\rho,p)$ together with 
$p(\rho,S)\to a^2(\rho,p)$ using the \eos/ relation \eqref{asq.rel}.
In particular, the variational derivatives transform as 
\begin{subequations}
\begin{align}
& \frac{\delta}{\delta U}\Big|_{(U,\rho,S)} = \frac{\delta}{\delta U}\Big|_{(U,\rho,p)},
\\
& \frac{\delta}{\delta \rho}\Big|_{(U,\rho,S)} = \frac{\delta}{\delta \rho}\Big|_{(U,\rho,p)} +a^2\frac{\delta}{\delta p}\Big|_{(U,\rho,p)}, 
\label{rho.der.rel}\\
& \frac{\delta}{\delta S}\Big|_{(U,\rho,S)} = (\partial p/\partial S)|_\rho \frac{\delta}{\delta p}\Big|_{(U,\rho,p)} . 
\end{align}
\end{subequations}
It is then straightforward to see that the Hamiltonian structure \eqref{Hamil.eqns} and \eqref{Hamil.op} 
becomes 
\begin{equation}
\partial_t 
\begin{pmatrix}
U \\ \rho \\ p
\end{pmatrix}
= \Hop_\gas 
\begin{pmatrix}
\delta H/\delta U \\ \delta H/\delta \rho \\ \delta H/\delta p 
\end{pmatrix}
\label{Hamil.eqns.gas}
\end{equation}
with 
\begin{equation}
\Hop_\gas =
\begin{pmatrix}
0 &  -D_r r^{1-n} &  r^{1-n}\frac{1}{\rho} p_r -\frac{1}{\rho}D_rr^{1-n}\rho a^2 \\
-r^{1-n} D_r & 0 & 0 \\
-r^{1-n}\frac{1}{\rho} p_r -r^{1-n}\rho a^2 D_r\frac{1}{\rho} & 0 & 0 
\end{pmatrix}
\label{Hamil.op.gas}
\end{equation}
where the same Hamiltonian \eqref{Hamil} is used, 
with the internal energy $e$ expressed in terms of $\rho$ and $p$. 
This expression can be obtained from 
$\frac{\partial e(\rho,S)}{\partial \rho} = \frac{\partial e(\rho,p)}{\partial \rho} + a^2 \frac{\partial e(\rho,p)}{\partial p} = p^2/\rho$, 
as shown by combining equation \eqref{e} and relation \eqref{rho.der.rel}. 

The associated Poisson bracket for radial gas dynamics is likewise obtained 
by putting 
$\Hop\to \Hop_\gas$ and $\nabla_{U,\rho,S} \to \nabla_{U,\rho,p}$
in the definition \eqref{PB}.

\subsection{Conserved integrals}

The results in \Ref{AncSeiDar} show that the kinematic conserved integrals 
for the radial Euler equations \eqref{U.eqn}--\eqref{S.eqn} 
consist of:
\begin{align}
\text{mass}\quad&
\frac{d}{dt}\int_{V(t)} \rho \,r^{n-1} dr 
= 0, 
\label{mass.integral}
\\
\text{total generalized-entropy}\quad&
\frac{d}{dt}\int_{V(t)} \rho  f(S) \,r^{n-1} dr 
= 0, 
\label{entropy.integral}
\\
\text{energy}\quad&
\frac{d}{dt}\int_{V(t)} \rho (\tfrac{1}{2} U^2 +e) \,r^{n-1} dr 
= - ( r^{n-1} p U )\Big|_{\partial V(t)}, 
\label{energy.integral}
\end{align}
in the case of a general \eos/ $p=p(\rho,S)$, 
where $e=\int p(\rho,S)/\rho^2\, d\rho$;
\begin{align}
& \text{dilational energy}\quad
\frac{d}{dt} \int_{V(t)} \big( t \rho (\tfrac{1}{2} U^2 +e) -\tfrac{1}{2} r \rho U \big) \,r^{n-1} dr
= - \big( r^{n-1} ( t  U -\tfrac{1}{2} r ) p \big)\Big|_{\partial V(t)}, 
\label{dilenergy.integral}
\\
& \text{similarity energy}\quad
\frac{d}{dt} \int_{V(t)} \big( t^2 \rho (\tfrac{1}{2} U^2 +e) -t\, r \rho U  +\tfrac{1}{2} r^2 \rho \big)\,r^{n-1} dr
= - \big( r^{n-1} t ( t  U - r ) p \big)\Big|_{\partial V(t)}, 
\label{simenergy.integral}
\end{align}  
in the case of a polytropic \eos/ $p=\k(S)\rho^{1+2/n}$,
where $e=\tfrac{n}{2}\k(S) \rho^{2/n}$
and $\k$ is an arbitrary function; 
\begin{equation}
\text{enthalpy flux}\quad
\frac{d}{dt} \int_{V(t)} U\, dr\Big|_\Esp 
= - ( e + p/\rho - \tfrac{1}{2} U^2 )\Big|_{\partial V(t)}
\label{enthaplyflux.integral}
\end{equation}
for a barotropic \eos/ $p=p(\rho)$,
where $e=\int p(\rho)/\rho^2\, d\rho$; 
\begin{equation}
\text{entropy-weighted energy}\quad
\frac{d}{dt} \int_{V(t)} \big( \tfrac{1}{2}\rho U^2 f(S) - K(S) \big)\, r^{n-1} dr\Big|_\Esp 
= - \big(\tfrac{1}{2} r^{n-1} U K(S) \big)\Big|_{\partial V(t)}
\label{nonisentropic.energy.integral}
\end{equation}
for an entropic \eos/ $p=\k(S)$, 
where $e=-\k(S)/\rho$ and $K(S)=\int f(S) \k'(S)\,dS$. 
Note $f$ denotes an arbitrary non-constant function of its argument. 

In these integrals, $V(t)$ is any radial domain that is transported in the flow, 
and each of the balance equations \eqref{mass.integral}--\eqref{nonisentropic.energy.integral}
holds on the solution space $\Esp$ of the radial fluid equations \eqref{U.eqn}--\eqref{eos}.
In general, a conserved integral on $V(t)$ is advected, namely, frozen into the flow, 
if the net flux through the moving boundary $\partial V(t)$ is zero. 
The mass \eqref{mass.integral} and total entropy \eqref{entropy.integral} 
are advected integrals,
whereas the other conserved integrals have non-zero net flux. 

A conserved integral is called kinematic if its density (modulo a total $r$-derivative) 
is a function of at most $t$, $r$, $\rho$, $S$, $U$; 
otherwise, if the density depends on $r$-derivatives of $\rho$, $S$, $U$, 
then it is called non-kinematic (or dynamical). 
The radial Euler equations \eqref{U.eqn}--\eqref{S.eqn} 
have been shown in \Ref{AncSeiDar} to possess two hierarchies of non-kinematic advected integrals. 

One hierarchy holds for a general \eos/ $p=p(\rho,S)$: 
\begin{equation}\label{integral.inv.Jhierarchy}
\I_l = \int_{V(t)} \rho f(J_0,J_1,\ldots,J_l) \,r^{n-1} dr, 
\quad
l=1,2,\ldots
\end{equation}
where 
\begin{equation}
J_l = \Rop^l S, 
\quad
l=0,1,2,\ldots
\label{J.hierarchy}
\end{equation}
are advected scalars given in terms of a recursion operator 
\begin{equation}
\Rop = (r^{1-n}/\rho) D_r . 
\label{inv.recursionop}
\end{equation}
The conserved integral \eqref{integral.inv.Jhierarchy} is 
non-trivial at order $l\geq1$ if and only if $f$ is nonlinear in its last argument, 
namely $f_{J_l J_l}\not\equiv 0$. 

Note that $\I_0$ is the entropy integral \eqref{entropy.integral}, 
since $J_0=S$. 

The other hierarchy holds only for an entropic \eos/ $p=\k(S)$: 
\begin{equation}\label{integral.inv.J1J2hierarchy}
\I'_l = \int_{V(t)} \rho f(J_0,J_1,J_{1,1},J_{2,1},\ldots,J_l,J_{1,l},J_{2,l}) \,r^{n-1} dr
\end{equation}
where 
\begin{align}
& J_{1,l}=\Rop^{l-1}\big( U^2 +\tfrac{2}{n}r p_r/\rho \big),
\quad
l=1,2,\ldots
\label{J1.hierarchy}
\\
& 
J_{2,l} = \Rop^{l-1}\big( A(r,U,p_r/\rho) - t \big)
= \int_0^1 \Rop^l\Big(r/\sqrt{U^2 +\tfrac{2}{n}(1-y^n) r p_r/\rho}\Big) dy ,
\quad
l=1,2,\ldots
\label{J2.hierarchy}
\end{align}
are advected scalars, 
with 
\begin{equation}
A(r,U,p_r/\rho) = \int_0^r \frac{dy}{\sqrt{U^2 + \tfrac{2}{n}(1-(y/r)^n)r p_r/\rho}}
= \int_0^1 \frac{r\,dy}{\sqrt{U^2 +\tfrac{2}{n}(1-y^n) r p_r/\rho}} . 
\label{A}
\end{equation}
(In each of the preceding integrals, $y$ denotes a different dummy variable.)
Here the conserved integral \eqref{integral.inv.J1J2hierarchy} 
is non-trivial at order $l\geq2$ if and only if $f$ is nonlinear in at least one of $J_{1,l}$ and $J_{2,l}$; 
at order $l=1$, it is non-trivial if and only if $f$ is non-constant in at least one of $J_{1,1}$ and $J_{2,1}$. 

Note that, as shown in \Ref{AncSeiDar},  
$\I'_1$ with $f=\tfrac{1}{2} J_{1,1}$ is equivalent (modulo a trivial integral) 
to the energy integral \eqref{energy.integral} with $p=\k(S) = -\rho e$.

\section{Point symmetries}\label{sec:pointsymms}

A Lie point symmetry of the radial Euler equations \eqref{U.eqn}--\eqref{S.eqn} 
is a one-parameter transformation group on $(t,r,\rho,S,U)$ 
with a generator
\begin{equation}\label{pointsymm}
\X = \tau\partial_t + \xi\partial_r + \eta^\rho\partial_\rho+ \eta^S\partial_S+ \eta^U\partial_U
\end{equation}
whose coefficients are functions of $(t,r,\rho,S,U)$ 
such that the solution space $\Esp$ of the equations is mapped into itself. 
The transformation group can be obtained from the generator via 
$(t,r,\rho,S,U)\to \exp(\epsilon\X)(t,r,\rho,S,U)$ 
where $\epsilon$ is the group parameter. 

The action of a symmetry transformation group on solutions $(U(t,r),\rho(t,r),S(t,r))$
is given by 
\begin{equation}
\begin{aligned}
U(t,r) & \to U(t,r) + \epsilon P^U|_{(U(t,r),\rho(t,r),S(t,r))} +O(\epsilon^2), \\
\rho(t,r) & \to \rho(t,r) + \epsilon P^\rho|_{(U(t,r),\rho(t,r),S(t,r))} +O(\epsilon^2), \\
S(t,r) & \to S(t,r) + \epsilon P^S|_{(U(t,r),\rho(t,r),S(t,r))} +O(\epsilon^2), 
\end{aligned}
\end{equation}
where 
\begin{equation}\label{canonical}
P^\rho = \eta^\rho -\tau\rho_t - \xi\rho_r,
\quad
P^U = \eta^U -\tau U_t - \xi U_r,
\quad
P^S = \eta^S -\tau S_t - \xi S_r.
\end{equation}
The infinitesimal form of this action is given by a generator 
\begin{equation}
\hat\X = P^\rho \partial_\rho+ P^S \partial_S + P^U \partial_U
\label{symm}
\end{equation}
which is called the characteristic form of the symmetry. 

The solution space $\Esp$ will be invariant if and only if the prolongation of $\X$ 
applied to the radial Euler equations \eqref{U.eqn}--\eqref{S.eqn} 
vanishes when evaluated on $\Esp$. 

A simpler, modern formulation of invariance \cite{Olv-book,BCA,Anc-review}
comes from using the characteristic form of generator
and is given by the condition that the prolongation of $\hat\X$ 
applied to the equations \eqref{U.eqn}--\eqref{S.eqn} (namely, their Frechet derivative) 
must vanish on $\Esp$: 
\begin{subequations}\label{symm.deteqns}
\begin{align}
& 
\big( D_t P^U + U D_r P^U +U_r P^U + D_r( p_S P^S + p_{\rho} P^\rho )/\rho - ( p_S S_r + p_{\rho} \rho_r )P^\rho /\rho^2 \big)\big|_\Esp =0, 
\\
& 
\big( D_t P^\rho + D_r (U P^\rho +\rho P^U) +\tfrac{n-1}{r} (U P^\rho +\rho P^U) \big)\big|_\Esp= 0, 
\\
&
\big( D_t P^S + U D_r P^S +S_r P^U \big)\big|_\Esp= 0 , 
\end{align}
\end{subequations}
where $D_t$ and $D_r$ respectively denote 
the total $t$-derivative and total $r$-derivative 
(see Appendix~\ref{app:identities}). 
These three determining equations split with respect to derivatives of $(U,\rho,S)$, 
yielding an overdetermined system of PDEs which can be solved
for $\tau$, $\xi$, $\eta^U$, $\eta^\rho$, $\eta^S$ along with $p(\rho,S)\neq$const.\ and $n\neq1$. 
This gives the following classification result. 
Remarks on the computation are given in Appendix~\ref{app:computation}. 

\begin{theorem}\label{thm:pointsymms}
(i) For a general \eos/ \eqref{eos}, 
the Lie point symmetries are generated by 
a time-translation $\partial_t$ and a dilation $t\partial_t + r\partial_r$. 
(ii) Additional Lie point symmetries exist only for the \esos/ 
shown in Table~\ref{table:symms} and their specializations, 
modulo an additive constant. 
(iii) A classification of all admitted maximal point-symmetry algebras is shown in Table~\ref{table:symmalgs}. 
\end{theorem}

\begin{table}[h]
\begin{tabular}{c|c}
\hline
$p$ 
& \# extra symmetries
\\
\hline
\hline
$\k(S)f(\rho)$
& 1
\\
\hline
$f(\rho) + \k(S)$ 
& 1
\\
\hline
$f(\k(S)\rho)\rho^{1+q}$,
$q\neq-1$
& 1
\\
\hline
$f(\k(S)\rho) + k\ln\rho$ 
& 1
\\
\hline
$f(\rho)$
& 1
\\
\hline
$\k(S)\rho^{1+2/n}$
& 1
\\
\hline
$\k(S)$
& 1
\\
\hline
\end{tabular}
\caption{Equations of state of maximal generality admitting extra Lie point symmetries. $f$ and $\k$ are non-constant functions.}
\label{table:symms}
\end{table}

The cases in Table~\ref{table:symms} are organized by generality of the \eos/. 
In each case, the number of extra symmetries counts 
only those symmetries that do not arise by a linear combination of 
symmetries inherited from intersections of more general cases. 

In Table~\ref{table:symmalgs}, 
the cases are organized by dimension 
and arise from specializations of the \esos/ listed in Table~\ref{table:symms}. 
The resulting classification is complete in the sense that
every point symmetry admitted for any given \eos/ appears 
among a linear combination of the listed symmetries or special cases of them. 
The notation for the algebras is taken from \Ref{PopBoyNesLuf}. 

\begin{table}[h]
\hbox{\hspace{-0.4in}
\begin{tabular}{c||c||c||c}
\hline
case
& $p$
& generators \& non-zero commutators
& algebra
\\
\hline
\hline
1 & $f(\rho,S)$
& $\X_1=\partial_t$, $\X_2=t\partial_t + r\partial_r$
& $\A{2,1}$
\\
& 
&
$[\X_1,\X_2] = \X_1$
&
\\
\hline
\hline
2  & $\k(S)f(\rho)$
& $\X_1$, $\X_2$, $\X_{\iii} =r\partial_r + U\partial_U + \frac{2\k(S)}{\k'(S)}\partial_S$
& $\A{2,1}\oplus\A{1}$
\\
& 
&
$[\X_1,\X_2] = \X_1$
&
\\
\hline
3 & $\k(S) + f(\rho)$
& $\X_1$, $\X_2$, $\X_{\vi} =\frac{1}{\k'(S)}\partial_S$
& $\A{2,1}\oplus\A{1}$
\\
& 
&
$[\X_1,\X_2] = \X_1$
&
\\
\hline
4 & $f(\k(S)\rho)\rho^{1+q}$
& $\X_1$, $\X_2$, $\X_{\ii} =qr\partial_r + qU\partial_U +2\rho\partial_\rho -\frac{2\k(S)}{\k'(S)}\partial_S$
& $\A{2,1}\oplus\A{1}$
\\
& $q\neq-1$
&
$[\X_1,\X_2] = \X_1$
&
\\
\hline
5 & $f(\k(S)\rho) + k\ln\rho$
& $\X_1$, $\X_2$, $\X_{\viii} =r\partial_r + U\partial_U -2\rho\partial_\rho +\frac{2\k(S)}{\k'(S)}\partial_S$
& $\A{2,1}\oplus\A{1}$
\\
& 
&
$[\X_1,\X_2] = \X_1$
&
\\
\hline
\hline
6  & $\k(S)\rho^{1+q}$
& $\X_1$, $\X_2$, $\X_{\iii}$, $\X_{\iv}=qr\partial_r + qU\partial_U +2\rho\partial_\rho$
& $\A{2,1}\oplus 2\A{1}$
\\
& $q\neq-1$
& 
$[\X_1,\X_2] = \X_1$
& 
\\
\hline
7 & $\k(S) + k\ln\rho$
& $\X_1$, $\X_2$, $\X_{\vi}$, $\X_{\vii} = r\partial_r + U\partial_U -2\rho\partial_\rho$
& $\A{2,1}\oplus 2\A{1}$
\\
& $k\neq 0$
&
$[\X_1,\X_2] = \X_1$
& 
\\
\hline
\hline
8 & $\k(S)\rho^{1+2/n}$
& $\X_1$, $\X_2$, $\X_{\iii}$, $\X_{\iv}'= r\partial_r + U\partial_U +n\rho\partial_\rho$,
& $\mathfrak{sl}(2,\Rnum)\oplus 2\A{1}$
\\
& 
& $\X_{\v} =t^2\partial_t + r\,t\partial_r + (r-tU)\partial_U -nt\rho\partial_\rho$
&
\\
&&
$[\X_1,\X_2]=\X_1$, $[\X_1,\X_{\v}]=2\X_2-\X'_{\iv}$, $[\X_2,\X_{\v}]=\X_{\v}$
&
\\
\hline
\hline
9 & $f(\rho)$
& $\X_1$, $\X_2$, $\X_{\ix} =F(S)\partial_S$
& $\A{2,1}\oplus\A{\infty}$
\\
& $(\rho f')'\neq 0$
&
$[\X_1,\X_2] = \X_1$
&
\\
& $(\rho f'/f)'\neq 0$ 
&&
\\
\hline
10 & $\k(S)$
& $\X_1$, $\X_2$, $\X_{\vii}$, 
& $\A{2,1}\oplus\A{1}\oplus\A{\infty}$
\\
& 
& $\X_{\vvi} =((F(S)\k'(S))'/\k'(S)) \rho\partial_\rho + F(S)\partial_S$
&
\\
&&
$[\X_1,\X_2] = \X_1$
&
\\
\hline
11 & $k\ln\rho$
& $\X_1$, $\X_2$, $\X_{\vii}$, $\X_{\ix}$
& $\A{2,1}\oplus \A{1}\oplus\A{\infty}$
\\
& $k\neq0$
&
$[\X_1,\X_2] = \X_1$
& 
\\
\hline
12 & $k\rho^{1+q}$
& $\X_1$, $\X_2$, $\X_{\iv}$, $\X_{\ix}$ 
& $\A{2,1}\oplus \A{1}\oplus\A{\infty}$
\\
& $k\neq0$, $q\neq-1$
&
$[\X_1,\X_2]=\X_1$
&
\\
\hline
13 & $k\rho^{1+2/n}$
& $\X_1$, $\X_2$, $\X_{\v}$, $\X_{\iv}'$, $\X_{\ix}$ 
& $\mathfrak{sl}(2,\Rnum)\oplus\A{1}\oplus\A{\infty}$
\\
& $k\neq0$
&
$[\X_1,\X_2]=\X_1$, $[\X_1,\X_{\v}]=2\X_2-\X'_{\iv}$, $[\X_2,\X_{\v}]=\X_{\v}$
&
\\
\hline
\hline
\end{tabular}
}
\caption{Maximal Lie point symmetry algebras. \newline
$f$ and $\k$ are non-constant functions; $F$ is a non-zero function.}
\label{table:symmalgs}
\end{table}

Note that the \esos/ in cases 6 and 7 in Table~\ref{table:symmalgs}
do not appear in Table~\ref{table:symms}. 
Their symmetries arise as linear combinations of the symmetries inherited from more general cases. 
Specifically, case 6 is given by the intersection of 
case 2 with $f=\rho^{1+q}$ and case 5 with $f=(\k(S)^{\frac{1}{q+1}}\rho)^{1+q}$, $k=0$, 
whereby 
$\X_{\iv}= (q+1)\X_{\iii}|_{f=\rho^{1+q}}-\X_{\viii}|_{f=(\k(S)^{\frac{1}{q+1}}\rho)^{1+q},k=0}$. 
Likewise, case 7 is given by the intersection of 
case 3 with $f=k\ln\rho$ 
and case 4 with $f=k\ln(e^{\k(S)/k}\rho)$, $q=-1$,
whereby 
$\X_{\vii}= {-2k} \X_{\vi}|_{f=k\ln\rho} -\X_{\ii}|_{f=k\ln(e^{\k(S)/k}\rho),q=-1}$.

Similarly, the \esos/ in cases 11, 12, and 13 
do not appear in Table~\ref{table:symms},
because all of the symmetries in these cases are directly inherited from 
the intersection of case 9 with cases 7, 6, 8, respectively. 

Also note that case 8 contains a specialization of case 6, 
where $\X_{\iv}'=\tfrac{1}{q}\X_{\iv}|_{q=\frac{2}{n}}$. 

In cases 8 and 13, the $\mathfrak{sl}(2,\Rnum)$ subalgebra is generated by 
$\{\X_1,\X_2 -\tfrac{1}{2}\X'_{\iv},\X_{\v}\}$, 
since $\X'_{\iv}$ commutes with $\X_1$ and $\X_{\v}$. 

\textbf{Remark}:
For each case in Table~\ref{table:symmalgs}, 
it is straightforward to derive a system of differential equations and inequations 
that involve only $p$ and $n$, whose general solution yields $p$. 
Such a characterization of cases is useful for determining which case 
contains a given \eos/, by checking which system the \eos/ satisfies. 

To conclude this discussion of Lie point symmetries, 
the transformation groups generated by each symmetry will now be presented. 

\begin{proposition}\label{prop:symm.groups}
The infinitesimal symmetries listed in Table~\ref{table:symmalgs}
generate the following transformation groups
(with $\epsilon$ being the group parameter): 
\begin{align}
&\begin{aligned}
X_1 :
t\to t+\epsilon;
\quad\text{time translation}
\end{aligned}
\label{symm1}
\\
&\begin{aligned}
X_2 :
t\to e^{\epsilon}t,
\quad
r\to e^{\epsilon}r;
\quad\text{dilation}
\end{aligned}
\label{symm2}
\\
&\begin{aligned}
X_{\ii} & : 
r\to e^{q\epsilon}r,
\quad
U\to e^{q\epsilon}U,
\quad
\rho\to e^{2\epsilon}\rho,
\quad
S\to \k^{-1}(e^{-2\epsilon}\k(S));
\\&\quad\text{combined scaling \& entropy shift}
\end{aligned}
\label{symm.ii}
\\
&\begin{aligned}
X_{\vii} &: 
r\to e^{\epsilon}r,
\quad
U\to e^{\epsilon}U,
\quad
\rho\to e^{-2\epsilon}\rho,
\quad
S\to \k^{-1}(e^{2\epsilon}\k(S));
\\&\quad\text{combined scaling \& entropy shift}
\end{aligned}
\label{symm.vii}
\\
&\begin{aligned}
X_{\iii} &: 
r\to e^{\epsilon}r,
\quad
U\to e^{\epsilon}U,
\quad
S\to \k^{-1}(e^{2\epsilon}\k(S));
\\&\quad\text{combined scaling \& entropy shift}
\end{aligned}
\label{symm4}
\\
&\begin{aligned}
X_{\iv} :
r\to e^{q\epsilon}r,
\quad
U\to e^{q\epsilon}U,
\quad
\rho\to e^{2\epsilon}\rho;
\quad&\text{scaling}
\end{aligned}
\label{symm.iv}
\\
&\begin{aligned}
X_{\v} &: 
t\to t/(1-\epsilon t),
\quad
r\to r/(1-\epsilon t),
\quad
U\to (1-\epsilon t)U +\epsilon r, 
\quad
\rho\to (1-\epsilon t)^n\rho ;
\\&\quad\text{conformal similarity}
\end{aligned}
\label{symm.v}
\\
&\begin{aligned}
X_{\vi} :
S\to \k^{-1}(\k(S) +\epsilon);
\quad&\text{entropy change}
\end{aligned}
\label{symm.vi}
\\
&\begin{aligned}
X_{\ix} :
S\to H^{-1}(H(S) +\epsilon);\quad&\text{entropy change}
\end{aligned}
\label{symm.ix}
\\
&\begin{aligned}
X_{\vvi} &:
\rho\to \rho \k'(H^{-1}(H(S) +\epsilon))F(H^{-1}(H(S) +\epsilon))/(\k'(S)H(S)),
\quad
S\to H^{-1}(H(S) +\epsilon);
\\&\quad\text{entropy change}
\end{aligned}
\label{symm.vvi}
\end{align}
where $H'(y)=1/F(y)$. 
\end{proposition}

Some remarks are worthwhile. 
An invariant of the groups \eqref{symm.ii} and \eqref{symm.vii} is $\k(S)\rho$. 
The group \eqref{symm.v} acts as a conformal scaling on $t$, $r$, $U-t/r$, $\rho$, 
with a scaling factor $1-\epsilon t$, 
where the scaling weight of $t$, $r$, $U-t/r$ is $1$, and the scaling weight of $\rho$ is $n$. 
The group \eqref{symm.vi} corresponds to $\k(S)\to \k(S) +\epsilon$. 
Likewise, the groups \eqref{symm.ix} and \eqref{symm.vvi} have $H(S)\to H(S) +\epsilon$,
while $\k'(S)F(S)/\rho$ is an invariant of the latter.

\section{Casimirs}\label{sec:casimirs}

A Casimir is a non-trivial functional $C=\int_0^\infty \Phi\, r^{n-1}dr$ such that 
its Poisson bracket \eqref{PB} with an arbitrary functional $\F$ vanishes: 
\begin{equation}\label{casimir}
\{C,\F\} = 0 
\end{equation}
(modulo a trivial functional). 
Existence of a Casimir indicates that the Poisson bracket is degenerate. 
If the density $\Phi$ has no explicit dependence on $t$, 
then $C$ is a conserved integral, 
as a consequence of relation \eqref{time.evol}. 
A symmetry characterization of Casimirs is stated in the next section. 

The notion of a Casimir can be generalized by considering functionals 
\begin{equation}
C=\int_{V(t)} \Phi\, r^{n-1}dr
\end{equation}
on a radial domain $V(t)$ that is transported by the fluid flow. 
From the definition \eqref{PB} of the Poisson bracket, 
the condition \eqref{casimir} is equivalent to $\Hop\nabla_{U,\rho,S} C = 0$,
where $\Hop$ is the underlying Hamiltonian (co-symplectic) operator. 
This latter condition can be used as a determining system 
to find the Casimirs of the radial Euler equations \eqref{U.eqn}--\eqref{S.eqn}. 
After simplification, the determining system is given by
\begin{equation}\label{casimir.detsys}
E_U(r^{n-1}\Phi) =0,
\quad
D_r(r^{1-n} E_\rho(r^{n-1}\Phi)) = r^{1-n}(S_r/\rho) E_S(r^{n-1}\Phi)
\end{equation}
in terms of the Euler operators $E_v$ (see Appendix~\ref{app:identities})
with respect to $v=(U,\rho,S)$. 
It is computationally straightforward to determine all Casimirs 
given by conserved densities $\Phi(r,U,\rho,S,U_r,\rho_r,S_r)$ up to first order, 
modulo trivial conserved densities. 

\begin{proposition}
For a general \eos/ for the radial Euler equations \eqref{U.eqn}--\eqref{S.eqn}, 
Casimirs having a conserved density up to first order 
are given by $\Phi = \rho f(J_0,J_1)$, 
where $J_0=S$ and $J_1=r^{1-n}S_r/\rho$ are the lowest-order invariants 
in the hierarchy \eqref{J.hierarchy}. 
No other first-order Casimirs exist for special \esos/. 
\end{proposition}

Since all first-order Casimirs $C=\int_{V(t)} \rho f(J_0,J_1)\,r^{n-1} dr$ 
coincide with the advected integrals $\I_1$ of order $1$, 
a natural question is whether the entire hierarchy of 
these integrals \eqref{integral.inv.Jhierarchy} are Casimirs. 

\begin{theorem}\label{thm:casmirs}
For a general \eos/ \eqref{eos}, 
every advected integral \eqref{integral.inv.Jhierarchy} is a Casimir. 
\end{theorem}
The proof will be given in the next subsection. 

In contrast, 
in the other hierarchy of advected integrals \eqref{integral.inv.J1J2hierarchy},
which hold only for entropic \esos/, 
none are Casimirs apart from the ones that also belong to the first hierarchy. 
This result is easily seen from the first equation in the determining system \eqref{casimir.detsys}, 
which implies that the conserved density $\Phi$ in a Casimir 
must have no essential dependence on $U$. 
Thus, all of the advected integrals involving at least one of $J_{1,l}$, $J_{2,l}$, $l=1,2,\ldots$, 
cannot be Casimirs since they depend explicitly on $U$ (and its $r$-derivatives). 

The question of whether there exist any additional Casimirs is a much harder problem 
which will be left for elsewhere.

\subsection{Proof of Theorem~\ref{thm:casmirs}}

The determining system \eqref{casimir.detsys} 
can be expressed in the form 
\begin{equation}
E_U(\tilde\Phi)=0,
\quad 
J_1 E_S(\tilde\Phi)= D_r E_{\tilde\rho}(\tilde\Phi)
\label{casimir.deteqns}
\end{equation}
where $\tilde\rho = r^{n-1}\rho$, 
and $\tilde\Phi = r^{n-1}\Phi$. 
Consider, hereafter, $\Phi=\rho f(J_l)$. 

The proof is by induction. 
Since $J_0= S$, 
then $\tilde\Phi= r^{n-1} \rho f(J_0)=\tilde\rho f(S)$, 
which is the density in the advected generalized-entropy integral \eqref{entropy.integral}. 
Hence, equations \eqref{casimir.deteqns} hold for $\tilde\Phi = \tilde\rho f(J_0)$. 

Now suppose that the determining equations \eqref{casimir.deteqns} 
hold for $\tilde\Phi = \tilde\rho f(J_k)$, $k\geq0$. 
The induction step requires showing that the determining equations 
then hold for $\tilde\Phi = \tilde\rho f(J_{k+1})$. 
This will be accomplished by splitting the equations with respect to derivatives of $f$,
which yields an equivalent system formulated in terms of Euler-Lagrange operators 
applied to the invariant $J_l$ as follows. 

\begin{lemma}\label{lem:split.casimir.sys}
The Casimir determining equations \eqref{casimir.deteqns}
are equivalent to the split system 
\begin{subequations}\label{Jn.sys}
\begin{align}
& J_1 E_S(J_k) = \Rop J_k + D_rE_{\tilde\rho}(J_k) ,
\\
& J_1 E^{(i)}_S(J_k) = D_r E^{(i)}_{\tilde\rho}(J_k) - E^{(i-1)}_{\tilde\rho}(J_k),
\quad
i=1,2,\ldots . 
\end{align}
\end{subequations}
\end{lemma}

To derive this system \eqref{Jn.sys}
from the two determining equations \eqref{casimir.deteqns}, 
observe that the first determining equation holds identically, 
since $J_k$ has no dependence on $U$ and its derivatives. 
Next, the second determining equation can be expressed in terms of derivatives of $f$ 
by use of the relation 
\begin{equation}
\begin{aligned}
J_1 \sum_{i\geq0} ({-D_r})^i(\tilde\rho f'(J_k)) E^{(i)}_S(J_k)
& = \tilde\rho f'(J_k) \big(\Rop J_k + D_rE_{\tilde\rho}(J_k)\big)
\\&\qquad
+ \sum_{i\geq1} ({-D_r})^i(\tilde\rho f'(J_k)) \big(D_r E^{(i)}_{\tilde\rho}(J_k) - E^{(i-1)}_{\tilde\rho}(J_k)\big) . 
\end{aligned}
\label{Jn.deteqn}
\end{equation}
(This relation holds by Euler-operator identity \eqref{Eop.rel1}, 
with $a=\tilde\rho$ and $b=J_k$.)
Because $f$ is an arbitrary function of $J_k$, 
the coefficients of  $({-D_r})^i(\tilde\rho f'(J_k))$ for $i=0,1,2,\ldots$ 
on each side of equation \eqref{Jn.deteqn} must be equal. 
As a result, this equation splits into the system \eqref{Jn.sys}. 

The proof of the induction step now starts from 
the determining equations \eqref{casimir.deteqns} with $\tilde\Phi = \tilde\rho f(J_{k+1})$,
where $J_{k+1}=\Rop J_k = (D_r J_k)/\tilde\rho$ 
via the recursion operator \eqref{inv.recursionop}. 
The first equation holds identically, since $J_k$ has no dependence on $U$ and its derivatives. 
The second equation splits similarly to the system \eqref{Jn.sys} 
by use of the relation 
\begin{equation}
\begin{aligned}
J_1 \sum_{i\geq0} ( ({-D_r})^{i+1} f'(J_{k+1}) ) E^{(i)}_S(J_k)
& = -D_r f'(J_{k+1})\big( J_{k+1} + D_r E_{\tilde\rho}(J_k) \big)
\\&\qquad
+ \sum_{i\geq1} ( ({-D_r})^{i+1} f'(J_{k+1}) ) \big(D_r E^{(i)}_{\tilde\rho}(J_k) - E^{(i-1)}_{\tilde\rho}(J_k)\big)
\end{aligned}
\end{equation}
(which holds by the Euler-operator identity \eqref{Eop.rel2},
with $a=\tilde\rho$ and $b=J_k$). 
It is easy to see that the coefficients of  $({-D_r})^{i+1} f'(J_{k+1}))$ for $i=0,1,2,\ldots$
on each side of this equation are equal due to the split system \eqref{Jn.sys}. 
Hence, the determining equations \eqref{casimir.deteqns} 
hold for $\tilde\Phi = \tilde\rho f(J_{k+1})$.
This establishes the induction.

The preceding argument can be extended straightforwardly to the general case
where $f$ depends on all invariants $J_0,J_1,\ldots, J_l$. 
This completes the proof of Theorem~\ref{thm:casmirs}.

\section{Hamiltonian symmetries}\label{sec:Hamilsymms}

A generalization of infinitesimal Lie point symmetries arises from 
allowing the components in the generator \eqref{pointsymm} 
to depend additionally on derivatives of $U$, $\rho$, $S$, 
such that the determining equations \eqref{symm.deteqns} hold 
using the characteristic form \eqref{symm} of the generator. 
If its components $(P^U,P^\rho,P^S)$ involve derivatives up to order $k\geq 1$, 
then such a generator is called a symmetry of order $k$, 
or sometimes, a generalized (higher-order) symmetry. 

Any generalized symmetry in characteristic form \eqref{symm} 
can be expressed in an equivalent form \eqref{pointsymm} 
where $\tau$ and $\xi$ are any functions of $t$, $r$, $U$, $\rho$, $S$, and their derivatives, 
while $\eta^U$, $\eta^\rho$, $\eta^S$ are given by the relations \eqref{canonical}
in terms of $(P^U,P^\rho,P^S)$. 
Lie point symmetries are characterized by the property that 
there is a unique choice of $\tau$ and $\xi$ depending only on $t$, $r$, $U$, $\rho$, $S$, 
for which $\eta^U$, $\eta^\rho$, $\eta^S$ have no dependence on derivatives of $U$, $\rho$, $S$. 

One main property of the Hamiltonian structure \eqref{Hamil.eqns}--\eqref{Hamil.op}
is that, for any conserved integral $\G=\int_{V(t)}\Phi\,r^{n-1} dr$, 
the action of the Hamiltonian operator $\Hop$
yields a generalized symmetry \eqref{symm} whose components are given by
\begin{equation}
\big( P^U ,\ P^\rho,\  P^S \big)^\t = \Hop \nabla_{U,\rho,S}\G, 
\end{equation}
namely
\begin{equation}\label{hamil.symm}
\begin{aligned}
P^U & = -D_r(r^{1-n} E_\rho(r^{n-1}\Phi)) +r^{1-n}(S_r/\rho) E_S(r^{n-1}\Phi),
\\
P^\rho & = -r^{1-n}D_r E_U(r^{n-1}\Phi),
\\
P^S & = -r^{1-n}(S_r/\rho) E_U(r^{n-1}\Phi). 
\end{aligned}
\end{equation}
A symmetry of this form is called a Hamiltonian symmetry. 

A general result in Hamiltonian theory \cite{Olv-book} states that 
if $\F$ and $\G$ are two conserved integrals, 
then the commutator of the corresponding Hamiltonian symmetries 
is a Hamiltonian symmetry corresponding to the Poisson bracket \eqref{PB} of $\F$ and $\G$. 
Thus, the Poisson bracket algebra of a closed set of conserved integrals 
is isomorphic to the Lie algebra of the corresponding set of Hamiltonian symmetries. 
If a conserved integral yields a Hamiltonian symmetry which is trivial, $\hat\X \equiv 0$, 
then it is a Casimir. 
In general, Hamiltonian symmetries may not exhaust all of the symmetries 
admitted by the Hamiltonian equations of motion. 

The Hamiltonian symmetries arising from all of the kinematic conserved integrals \eqref{mass.integral}--\eqref{nonisentropic.energy.integral}
possessed by the radial Euler equations \eqref{U.eqn}--\eqref{S.eqn} 
will now be derived from expressions \eqref{hamil.symm}. 

The mass and generalized-entropy integrals \eqref{mass.integral}--\eqref{entropy.integral} 
yield a trivial symmetry $\hat\X=0$, 
since they are special cases of the Casimir $C=\I_0=\int_{V(t)} \rho f(J_0)\,r^{n-1} dr$ 
with $f$ being an arbitrary function of $J_0=S$. 

For the remaining 5 kinematics conserved integrals \eqref{energy.integral}--\eqref{nonisentropic.energy.integral}, 
the Hamiltonian symmetries are shown in Table~\ref{table:kin.symms}.
For each one, 
a suitable choice of $\tau$ and $\xi$ is made so that the symmetry generator takes 
the simplest possible form. 
The respective choices are given by the coefficients of $D_t$ and $D_r$ 
in the expressions for $(P^U,P^\rho,P^S)$ in terms of $(U,\rho,S)$. 

As expected, 
the energy integral \eqref{energy.integral} corresponds to time-translation symmetry. 
The dilational and similarity energies \eqref{dilenergy.integral} and \eqref{simenergy.integral} 
correspond to scaling and conformal similarity symmetries, 
which are Lie point symmetries. 
The resulting transformation groups generated by these symmetries
are shown in section~\ref{sec:pointsymms}. 

In contrast, 
both the enthalpy flux integral \eqref{enthaplyflux.integral} 
and the entropy-weighted energy integral \eqref{nonisentropic.energy.integral}
correspond to first-order generalized symmetries. 
Each of these symmetries generates a transformation group acting on solutions of
the radial Euler equations \eqref{U.eqn}--\eqref{S.eqn},
where the group is defined by the system of first-order PDEs
for $(U^*(t,r;\epsilon),\rho^*(t,r;\epsilon),S^*(t,r;\epsilon))$,
which consists of  
\begin{equation}\label{1stordsymm.deteqns}
U^*_\epsilon = P^U|_{(U^*,\rho^*,S^*)},
\quad
\rho^*_\epsilon = P^\rho|_{(U^*,\rho^*,S^*)},
\quad
S^*_\epsilon = P^S|_{(U^*,\rho^*,S^*)}
\end{equation}
in terms of the components of the symmetry generator,
with $\epsilon$ denoting the group parameter
such that $(U^*(t,r;0),\rho^*(t,r;0),S^*(t,r;0))=(U(t,r),\rho(t,r),S(t,r))$. 

For the symmetry arising from the enthalpy flux integral \eqref{enthaplyflux.integral}, 
the determining system \eqref{1stordsymm.deteqns} is given by 
\begin{equation}
U^*_\epsilon = 0, 
\quad
\rho^*_\epsilon = 0, 
\quad
S^*_\epsilon + r^{1-n} S^*_r/\rho^* = 0 . 
\end{equation}
Explicit integration of this system yields the transformation group 
\begin{equation}
U^*=U(t,r),
\quad
\rho^*=\rho(t,r),
\quad
S^*=S(M^{-1}(t,M(t,r)-\epsilon))
\end{equation}
where the function $M(r) = \int_0^r \rho(t,r)\, r^{n-1} dr$ 
is the mass contained in the radial domain $[0,r]$ at any fixed time $t$, 
and $M^{-1}$ denotes the inverse function. 
Similarly, 
the determining system \eqref{1stordsymm.deteqns} defined by 
the symmetry arising from the entropy-weighted energy integral \eqref{nonisentropic.energy.integral}
consists of 
\begin{equation}
U^*_\epsilon +f(S^*) U^*_t = 0, 
\quad
\rho^*_\epsilon +f(S^*) \rho^*_t -f'(S^*) S^*_r\rho^* = 0, 
\quad
S^*_\epsilon +f(S^*) S^*_t = 0 . 
\end{equation}
This first-order system can be integrated to obtain the transformation group 
given by 
\begin{equation}
U^*=U(\sigma,r),
\quad
\rho^*=\rho(\sigma,r) \big(1+\epsilon S_t(\sigma,r) f'(S^*)\big)^{S_r(\sigma,r)/S_t(\sigma,r)},
\end{equation}
with $S^*=S^*(t,r;\epsilon)$ being the implicit solution of 
\begin{equation}
S^*=S(\sigma,r),
\quad
\sigma = t-\epsilon f(S^*) . 
\end{equation}

\begin{table}[h]
\hbox{\hspace{-0.5in}
\begin{tabular}{c|c||c||c}
\hline
conserved integral
& 
$p$
& 
$(P^U,P^\rho,P^S)$
& 
symmetry 
\\
\hline
energy \eqref{energy.integral}
& $f(\rho,S)$
& $-D_t(U,\rho,S)$
& time-translation $\X_1$
\\
\hline
dilational 
& $\k(S)\rho^{1+q}$
& $(U,\tfrac{2}{q}\rho,0)- rD_r(U,\rho,S)$
& scaling $\X_{\iv}$
\\
energy \eqref{dilenergy.integral}
&&&\\
\hline
similarity 
& $\k(S)\rho^{1+q}$
& $(r-tU, -nt\rho,0) -(t^2D_t +rt D_r)(U,\rho,S)$
& conformal similarity $\X_{\v}$
\\
energy \eqref{simenergy.integral}
&&&\\
\hline
enthalpy 
& $f(\rho)$
& $(0,0,{-J_1})$
& 1st-order $\X ={-J_1} \partial_S$
\\
flux \eqref{enthaplyflux.integral}
&&&\\
\hline
entropy-weighted 
& $\k(S)$
& $-f(S)D_t(U,\rho,S) +(0,f'(S)S_r\rho,0)$
& 1st-order 
\\
energy \eqref{nonisentropic.energy.integral}
&&&
$\X = f(S)\partial_t +f'(S)S_r\rho\partial_\rho$
\\
\hline
\end{tabular}
}
\caption{Hamiltonian symmetries from kinematic conserved integrals}
\label{table:kin.symms}
\end{table}

In addition to kinematic conserved integrals, 
the hierarchy of non-kinematic advected integrals \eqref{integral.inv.J1J2hierarchy}
for entropic \esos/ 
yield non-trivial Hamiltonian symmetries since, as shown in the previous section, 
none of them are Casimirs. 
The resulting symmetries are more complicated in comparison to the symmetries in Table~\ref{table:kin.symms}. 

Firstly, 
the two simplest advected integrals will be considered:
\begin{align}
& \I'_1\big|_{f=J_{1,1}} = \int_{V(t)} \rho J_{1,1}\, r^{n-1} dr , 
\label{J1.integral}
\\
& \I'_1\big|_{f=J_{2,1}} = \int_{V(t)} \rho J_{2,1}\, r^{n-1} dr , 
\label{J2.integral}
\end{align}
which involve the advected scalars  
$J_{1,1}=U^2 +\tfrac{2}{n}r p_r/\rho$ 
and $J_{2,1} = A(r,U,p_r/\rho) - t$ with $A(r,U,p_r/\rho)$ given by expression \eqref{A}. 

Recall that, as remarked in section~\ref{sec:eqns}, 
the advected integral \eqref{J1.integral} is equivalent to the energy integral \eqref{energy.integral} 
(modulo a trivial conserved integral)
specialized to the case of an entropic \eos/. 
For this conserved integral, 
the correspondence \eqref{hamil.symm} yields 
$(P^U,P^\rho,P^S)= 2 D_t(U,\rho,S)$ 
which is equivalent to a time-translation symmetry 
\begin{equation}\label{J1.symm}
\X_{J_{1,1}} = -2\X_1 .
\end{equation}

For the other advected integral \eqref{J2.integral},
the correspondence \eqref{hamil.symm} gives
$(P^U,P^\rho,P^S)= 
-(J_{1,2})_U D_r(U,\rho,S) -\big( (J_{1,2})_r,\rho(D_r (J_{1,2})_U+\tfrac{n-1}{r}(J_{1,2})_U),0 \big)$. 
This yields the 2nd-order symmetry
\begin{equation}\label{J2.symm}
\X_{J_{2,1}}=A_U \partial_r - A_r\partial_U -\rho(D_r A_U+\tfrac{n-1}{r}A_U)\partial_\rho . 
\end{equation}

The two symmetries \eqref{J1.symm} and \eqref{J2.symm} commute, namely
$[\pr\hat\X_{J_{1,1}}, \pr\hat\X_{J_{2,1}}]=0$
using their characteristic form \eqref{symm},
where $\pr$ denotes prolongation. 

Similar results hold for the whole hierarchy of advected integrals \eqref{integral.inv.J1J2hierarchy}. 
A proof is provided in the next subsection. 

\begin{theorem}\label{thm:J1J2.hamilsymms}
The advected integral 
\begin{equation}\label{fJ1J2.integrals}
\I'_l\big|_{f(J_{1,l},J_{2,l})} = \int_{V(t)} \rho f(J_{1,l},J_{2,l})\, r^{n-1} dr,
\end{equation}
for any $l\geq1$, 
yields the $l$th-order Hamiltonian symmetry 
\begin{equation}\label{fJ1J2.symm}
\X_{f(J_{1,l},J_{2,l})} = -2f_{1,l}^{(l-1)}\partial_t +A_U f_{2,l}^{(l-1)}\partial_r 
-A_r f_{2,l}^{(l-1)}\partial_U + \big( 2 D_t f_{1,l}^{(l-1)} -D_r(A_U f_{2,l}^{(l-1)}) -\tfrac{n-1}{r} A_U f_{2,l}^{(l-1)} \big)\rho\partial_\rho
\end{equation}
where $f_{\cdot,l}^{(i)}=(-\Rop)^{i} f_{J_{\cdot,l}}$, 
with $\Rop$ being the recursion operator \eqref{inv.recursionop}. 
\end{theorem}

The set of all symmetries \eqref{fJ1J2.symm} comprises a Lie algebra 
which has a non-trivial commutator structure. 
As an illustration, consider the lowest-order case $l=1$. 
Then a direct computation shows that 
\begin{equation}
[\pr\hat\X_{f(J_{1,1},J_{2,1})},\pr\hat\X_{g(J_{1,1},J_{2,1})}]
= \pr\hat\X_{h(J_{1,1},J_{2,1})},
\quad
h=2f_{J_{1,1}}g_{J_{2,1}} - 2f_{J_{2,1}}g_{J_{1,1}} . 
\end{equation}
The symmetry $\X_{h(J_{1,1},J_{2,1})}$ here will be non-trivial 
if and only if $h$ is not linear in $(J_{1,1},J_{2,1})$. 
For instance, if $f$ and $g$ are polynomials, 
then at least one of them must be at least quadratic,
otherwise $\X_{h(J_{1,1},J_{2,1})}$ will vanish. 
A similar result holds for $l\geq 2$. 

Thus, the radial Euler equations \eqref{U.eqn}--\eqref{S.eqn} possess 
a rich structure of Hamiltonian symmetries.

\subsection{Proof of Theorem~\ref{thm:J1J2.hamilsymms}}

The first step will use the following result
for evaluating the Euler operator applied to the densities
in the two integrals \eqref{fJ1J2.integrals}. 

\begin{lemma}
Let $K=K(r,U,\rho,S,S_r)$ and $f(K)$ 
be arbitrary (smooth) functions of their arguments. 
Denote $K_{i} = \Rop^i K$ and $f_K^{(i)}=(-\Rop)^i f'(K)$, $i\geq0$, 
using the recursion operator \eqref{inv.recursionop}. 
Then: 
\begin{align}
E_U(r^{n-1}\rho f(K_l)) &
= f_K^{(l)} E_U(r^{n-1}\rho K) , 
\label{Eop.U.K}\\
E_S(r^{n-1}\rho f(K_l)) &
= f_K^{(l)} E_S(r^{n-1}\rho K)  - (D_r f_K^{(l)}) E^{(1)}_S(r^{n-1}\rho K) , 
\label{Eop.S.K}\\
E_\rho(r^{n-1}\rho f(K_l)) &
= f_K^{(l)} E_\rho(r^{n-1}\rho K) 
+r^{n-1}\big( f - \Dop_l f' \big) , 
\label{Eop.rho.K}
\end{align}
where $\Dop_l = \sum_{0\leq i\leq l} K_{l-i} (-\Rop)^i$. 
Moreover: 
\begin{equation}
D_r\big( f(K_l) - \Dop_l f'(K_l) \big) = -K D_r f_K^{(l)} .
\label{Dop.rel}
\end{equation}
\end{lemma}
These identities \eqref{Eop.U.K}--\eqref{Eop.rho.K}
can be derived by the following descent argument. 
Firstly, consider the lefthand side of \eqref{Eop.U.K},
and successively apply the Euler-operator identities \eqref{Eop.rel2} and \eqref{Eop.rel3} 
using $v=U$, $a=r^{n-1}\rho$ and $b=K_l$.
This yields:
$E_U(r^{n-1}\rho f(K_l)) 
=\sum_{1\leq i\leq l} E^{(i-1)}_U(K_{l-1}) (-D_r)^{i} f'
=-\sum_{1\leq i\leq l-1} E^{(i-1)}_U(K_{l-2}) (-D_r)^{i} f_K^{(1)}$.
Iteration of this step leads to the righthand side of \eqref{Eop.U.K},
using the property that $E^{i}_U(K)=0$ for $i\geq1$ 
since $K$ does not depend on derivatives of $U$. 
Secondly, consider the lefthand side of \eqref{Eop.S.K}. 
The previous steps with $v=\rho$ lead to the righthand side of \eqref{Eop.S.K}
where the additional term arises because $K$ depends on $S_r$ 
(but not any higher derivatives of $S$). 
Thirdly, the derivation of \eqref{Eop.rho.K} is similar 
and uses properties that $E^{i}_\rho(K)=0$ and $E^{i}_\rho(a)=0$ for $i\geq1$,
as well as $E_\rho(a)=r^{n-1}$. 

Now, returning to the main proof, 
the respective cases $f=f(J_{1,1})$ and $f=f(J_{2,1})$ will be considered first. 
Both cases involve the same steps. 

\textbf{Case} $f=f(J_{1,1})$:
Put $K=J_{1,1}$ into the identities \eqref{Eop.U.K}--\eqref{Eop.rho.K}
where $r^{n-1}\rho K = r^{n-1}\rho U^2 + \tfrac{2}{n} (r/\rho) p_r$.
Use of 
\begin{equation}
\begin{gathered}
E_U(r^{n-1}\rho J_{1,1}) = 2 r^{n-1}\rho U,
\quad
E_\rho(r^{n-1}\rho J_{1,1}) = 2 r^{n-1} U^2,
\\
E_S(r^{n-1}\rho J_{1,1}) = -2 r^{n-1}p',
\quad
E^{(1)}_S(r^{n-1}\rho J_{1,1}) = \tfrac{2}{n} r^n p'
\end{gathered}
\end{equation}
leads to the expressions 
\begin{align}
& Q^U_{1,l} \equiv E_U(r^{n-1}\rho f(J_{1,l}))
= 2 r^{n-1}\rho U f_{J_{1,l}}^{(l)} , 
\label{QU.J1}\\
& Q^\rho_{1,l} \equiv E_\rho(r^{n-1}\rho f(J_{1,l})) 
= r^{n-1}\big( U^2 f_{J_{1,l}}^{(l)}   + f - \sum_{0\leq i\leq l} J_{1,l-i} f_{J_{1,l}}^{(i)} \big) , 
\label{Qrho.J1}\\
& Q^S_{1,l} \equiv  E_S(r^{n-1}\rho f(J_{1,l})) 
= -\tfrac{2}{n} p' D_r ( r^n f_{J_{1,l}}^{(l)} ) . 
\label{QS.J1}
\end{align}

The next step in the proof of this case 
is to substitute expressions \eqref{QU.J1}--\eqref{QS.J1}
into the correspondence \eqref{hamil.symm} to obtain 
the components $(P^U,P^\rho,P^S)$ of the Hamiltonian symmetry. 
Proceeding in order of simplicity:
first, 
\begin{equation}\label{PS.J1}
P^S = -(r^{1-n}S_r/\rho) Q^U_{1,l}
= -2 U S_r f_{J_{1,l}}^{(l)} ; 
\end{equation}
second, 
\begin{equation}\label{Prho.J1}
P^\rho = - r^{1-n} D_r Q^U_{1,l}
= -2 ((\rho U)_r + \tfrac{n-1}{r}\rho U) f_{J_{1,l}}^{(l)} -2 \rho U D_r f_{J_{1,l}}^{(l)} ;
\end{equation}
and last, 
\begin{equation}\label{PU.J1}
\begin{aligned}
P^U & = (r^{1-n}S_r/\rho) Q^S_{1,l} - D_r(r^{1-n} Q^\rho_{1,l}) \\
& = -\tfrac{2}{n} (r^{1-n} p_r/\rho) D_r( r^n f_{J_{1,l} }^{(l)} )
- D_r( U^2 f_{J_{1,l}}^{(l)} + f -\Dop_l f'(J_{1,l}) ) \\
& = -2( U U_r +p_r/\rho ) f_{J_{1,l}}^{(l)} 
\end{aligned}
\end{equation}
which uses $D_r( f -\Dop_l f'(J_{1,l}) ) = -J_{1,l} D_r f_{J_{1,l}}^{(l)}$ 
via identity \eqref{Dop.rel}. 

For the final step in the proof of the first case, 
the $r$-derivatives in the expressions \eqref{PS.J1}--\eqref{PU.J1}
can replaced in terms of $t$-derivatives through 
the radial Euler equations \eqref{U.eqn}--\eqref{S.eqn}. 
Likewise, since $J_{1,l}$ is an advected invariant,
it satisfies $D_t J_{1,l} + U D_r J_{1,l} =0$,
which implies $U D_r f_K^{(l)}(J_{1,l})  = -D_t f_K^{(l)}(J_{1,l})$
since $\Rop$ is a recursion operator on advected invariants. 
Hence, this yields 
\begin{equation}\label{Ps.J1}
P^S = 2 S_t f_{J_{1,l}}^{(l)},
\quad
P^\rho = 2 D_t( \rho f_{J_{1,l}}^{(l)} ),
\quad
P^U = 2 U_t f_{J_{1,l}}^{(l)},  
\end{equation}
which can be expressed more simply as
\begin{equation}
(P^U,P^\rho,P^S) = 2 f_{J_{1,l}}^{(l)} D_t(U,\rho,S) + 2(0,\rho D_t f_{J_{1,l}}^{(l)},0) , 
\end{equation}
giving the Hamiltonian symmetry
\begin{equation}
\X_{f(J_{1,l})} = -2f_{1,l}^{(l-1)}\partial_t + ( 2 D_t f_{1,l}^{(l-1)} )\rho\partial_\rho .
\end{equation}

\textbf{Case} $f=f(J_{2,1})$:
Put $K=J_{2,1}$ into the identities \eqref{Eop.U.K}--\eqref{Eop.rho.K}
where $r^{n-1}\rho K = r^{n-1}\rho (A(r,U,p_r/\rho)-t)$. 
Use of 
\begin{equation}
\begin{gathered}
E_U(r^{n-1}\rho J_{2,1}) = r^{n-1}\rho A_U,
\quad
E_\rho(r^{n-1}\rho J_{2,1}) = r^{n-1}( J_{2,1} - p_r A_{p_r} ), 
\\
E_S(r^{n-1}\rho J_{2,1}) = -\k'(S) D_r( r^{n-1}\rho A_{p_r} ), 
\quad
E^{(1)}_S(r^{n-1}\rho J_{2,1}) = r^{n-1}\rho \k'(S) A_{p_r} 
\end{gathered}
\end{equation}
leads to the expressions 
\begin{align}
& Q^U_{2,l} \equiv E_U(r^{n-1}\rho f(J_{2,l}))
= r^{n-1}\rho A_U f_{J_{2,l}}^{(l)},
\label{QU.J2}\\
& Q^\rho_{2,l} \equiv E_\rho(r^{n-1}\rho f(J_{2,l})) 
= r^{n-1}( f - p_r A_{p_r} f_{J_{2,l}}^{(l)} - \sum_{1\leq i\leq l} J_{2,l-i} f_{J_{2,l}}^{(i)}  \big), 
\label{Qrho.J2}\\
& Q^S_{2,l} \equiv  E_S(r^{n-1}\rho f(J_{2,l})) 
= - \k'(S) D_r( r^{n-1}\rho A_{p_r} f_{J_{2,l}}^{(l)} ) . 
\label{QS.J2}
\end{align}
The components $(P^U,P^\rho,P^S)$ of the Hamiltonian symmetry are then given by 
\begin{align}
P^S & = -(r^{1-n}S_r/\rho) Q^U_{2,l}
= - S_r A_U f_{J_{2,l}}^{(l)} , 
\label{PS.J2}
\\
P^\rho & = - r^{1-n} D_r Q^U_{1,l}
= -D_r( \rho A_U f_{J_{2,l}}^{(l)} ) - \tfrac{n-1}{r}\rho A_U f_{J_{2,l}}^{(l)} , 
\label{Prho.J2}
\\
P^U & = (r^{1-n}S_r/\rho) Q^S_{1,l} - D_r(r^{1-n} Q^\rho_{1,l}) 
= -( A_r +U_r A_U ) f_{J_{2,l}}^{(l)} 
\label{PU.J2}
\end{align}
similarly to the previous case. 
Hence, 
\begin{equation}
(P^U,P^\rho,P^S) = - A_U f_{J_{2,l}}^{(l)} D_r(U,\rho,S) 
- \big( A_r f_{J_{2,l}}^{(l)},(D_r(A_U f_{J_{2,l}}^{(l)}) + \tfrac{n-1}{r} A_U)\rho, 0 \big)
\end{equation}
gives the Hamiltonian symmetry
\begin{equation}
\X_{f(J_{2,l})} = A_U f_{2,l}^{(l-1)}\partial_r -A_r f_{2,l}^{(l-1)}\partial_U 
-(D_r(A_U f_{2,l}^{(l-1)}) +\tfrac{n-1}{r} A_U f_{2,l}^{(l-1)} )\rho\partial_\rho .
\end{equation}

Finally, the general case $f=f(J_{1,l},J_{2,l})$ 
stated in Theorem~\ref{thm:J1J2.hamilsymms}
is obtained by the same steps. 

It is worth remarking is that the expressions 
\eqref{QU.J1}--\eqref{QS.J1} and \eqref{QU.J2}--\eqref{QS.J2} 
are the respective multipliers that correspond to the conserved integrals 
$\I'_l\big|_{f(J_{1,l})} = \int_{V(t)} \rho f(J_{1,l})\, r^{n-1} dr$
and
$\I'_l\big|_{f(J_{2,l})} = \int_{V(t)} \rho f(J_{2,l})\, r^{n-1} dr$, 
as shown by general results for conserved integrals of evolution equations
(see e.g. \cite{Olv-book,BCA,Anc-review}).

\section{Concluding remarks}\label{sec:remarks} 

Radial fluid flow in $n>1$ dimensions possesses, unexpectedly, 
a rich structure of point symmetries and generalized symmetries. 
The point symmetries comprise 
time-translation, space-time dilation, scaling, conformal similarity, 
which are well known for non-radial flow in three dimensions \cite{Ibr-CRC}, 
as well as an entropy shift combined with various scalings, 
and an entropy change. 
These symmetries are found to hold for several different types of \esos/. 

The time-translation, scaling, and conformal similarity
are also Hamiltonian symmetries,
which arise from the kinematic conserved integrals 
for energy, dilational energy, and similarity energy, respectively. 
The two ``hidden'' kinematic conserved integrals, 
describing an enthalpy-flux quantity \eqref{enthaplyflux.integral} 
which holds for barotropic \esos/,
and an entropy-weighted energy \eqref{nonisentropic.energy.integral}
which holds for entropic \esos/, 
give rise to first-order generalized symmetries. 
Each of these symmetries are shown to produce a transformation group
acting on solutions of the equations for radial fluid flow. 

The hierarchy of advected conserved integrals
holding for a general \eos/ are proved to be Hamiltonian Casimirs,
which correspond to trivial symmetries. 
In contrast, the additional hierarchy of advected conserved integrals that hold 
only for an entropic \eos/ 
give rise to a corresponding hierarchy of non-trivial generalized symmetries. 
These symmetries are not inherited from any symmetries of 
$n$-dimensional non-radial fluid flow. 
The first-order generalized symmetries are explicitly shown to 
comprise a non-abelian Lie algebra. 

All of the preceding results carry over to radial gas dynamics through the well-known 
equivalence between the respective governing equations of 
$n$-dimensional gas dynamics and $n$-dimensional compressible fluid flow. 
Specifically, 
when a symmetry generator is expressed solely in terms of $t$, $r$, $U$, $\rho$, $p$, $e$, 
then it manifestly holds for both radial fluid flow and radial gas dynamics. 

There are several directions of interest for future investigation. 

One open question is to find a proof of whether the Casimir hierarchy comprises all Casimirs 
and likewise whether any more advected conserved integrals exist 
beyond the Casimirs and the ones in the hierarchy holding for an entropic \eos/. 
A similar question is whether any additional higher-order symmetries exist
other than the Hamiltonian ones. 
Related to this, an important problem is whether the radial compressible flow equations 
constitute an integral system with a bi-Hamiltonian structure, 
similarly to what occurs in $n=1$ dimensions \cite{Ver}. 

Another direction would be to generalize the results 
in the present paper and the preceding work in \Ref{AncSeiDar} to 
spherically-symmetric flows in two dimensions, 
which are more general than just purely radial flows. 

Looking towards applications, an interesting problem is to investigate what implications
the existence of the new conserved integrals will have for understanding 
the development and properties of finite-time singularities which are well known to occur 
in the Cauchy problem for classical solutions of one-dimensional compressible flows.

\appendix

\section{Computational remarks for symmetries and Casimirs}\label{app:computation}

The overdetermined system that arises from splitting the determining equations \eqref{symm.deteqns}
for $\tau,\xi,\eta^U,\eta^\rho,\eta^S$ along with $p(\rho,S)\neq$const.\ and $n\neq1$
contains 25 PDEs. 
Solving this system is a nonlinear problem 
which leads to many case distinctions in the solution process.
To obtain all solutions, it is important that no cases are lost 
when integrability conditions are used and when differential equations are integrated. 
The program CRACK \cite{Wol} is able reliably to carry out the computation,
where the overdetermined system is obtained by the program LiePDE. 

CRACK splits the computation repeatedly into cases whenever
equations factorize, or coefficients of functions that are to be substituted may be zero, 
or integration of single differential equations with parameters 
has more than one solution branch. 

Once all equations in every case have been solved, 
the solutions need to be merged into a complete case tree 
by eliminating solution cases that are contained in more general cases. 
In particular, CRACK may perform case splittings that are necessary to
complete the computation automatically but that do not provide new symmetries or new \esos/. 

It is straightforward to use LiePDE to determine 
if a case distinction in CRACK leads to a new solution case or not. 
The process consists of re-running CRACK from a call to LiePDE
with all of the free constants and free functions in $p(\rho,S)$ 
being taken as fixed (namely, they are not to be solved for) in the input, 
and thereby solving only for the symmetry components \eqref{canonical}. 
If the output contains fewer symmetries compared to the original solution case, 
then the case distinction that does not produce new symmetries 
and therefore is not necessary. 

LiePDE and CRACK have several strengths that are relevant 
in the present computations of Lie point symmetries.  
No cases are lost; the worst that happens is that some consistent set of equations are left unsolved, and this occurs in only one case. 
Nearly all steps are done fully automatically.
Any part of the computation can be done interactively if it is desired or needed. 
The whole computation of all cases with all integrations runs in a few seconds 
on a desktop computer. 

The results were checked independently by using Maple 
similarly to the computation of conserved integrals in \Ref{AncSeiDar}.

\section{Euler operator identities}\label{app:identities}

For a dependent variable $v$ and an independent variable $z$, 
the coordinate space $(z,v,v_z,v_{zz},\ldots)$ defines the jet space of $v(z)$. 
A total derivative with respect to $z$ is represented by the operator 
\begin{equation}
D_z = \partial_z + \sum_{j\geq 0} \partial_z^{j+1}v \partial_{\partial_z^j v} . 
\end{equation}
Then 
\begin{equation}
E_v = \sum_{j\geq 0} ({-D_z})^j \partial_{\partial_z^j v},
\quad
E^{(i)}_v = \sum_{j\geq 0} {\textstyle\binom{i+j}{i}} ({-D_z})^j \partial_{\partial_z^{i+j}v},
i=1,2,\ldots
\end{equation}
are, respectively, the Euler operator and the higher Euler operators
\cite{Olv-book,Anc-review}
with respect to $v$. 
Note that $E^{(0)}_v = E_v$. 

Let $a$ and $b$ be arbitrary smooth functions on the jet space of $v(z)$,
and let $f$ denote a smooth function of its argument. 

The following three identities 
can be derived similarly to the product rule for the Euler operator: 
\begin{equation}
E_v(af(b)) = \sum_{i\geq0} E^{(i)}_v(b) ({-D_z})^i(a f'(b)) + E^{(i)}_v(a)({-D_z})^i f(b)
\label{Eop.rel1} 
\end{equation}
\begin{equation}
E_v(af(b_{+1})) = \sum_{i\geq0} E^{(i)}_v(b)  ({-D_z})^{i+1} f'(b_{+1})
+ E^{(i)}_v(a) ({-D_z})^i(f(b_{+1})-b_{+1}f'(b_{+1})) 
\label{Eop.rel2} 
\end{equation}
\begin{equation}
\begin{aligned}
\sum_{i\geq0} E^{(i)}_v(b_{+1}) ({-D_z})^{i+1} f(b_{+1}) =& 
\sum_{i\geq0} E^{(i+1)}_v(b) ({-D_z})^{i+1}((D_z f'(b{+1}))/a)
\\&
-\sum_{i\geq0}\sum_{j\geq 0} {\textstyle\binom{i+j}{j}} E^{(i+j)}_v(a) (({-D_z})^j b_{+1})({-D_z})^{i}((D_z f'(b_{+1}))/a) 
\end{aligned}
\label{Eop.rel3}
\end{equation}
where $b_{+1} = (D_z b)/a$.

\section*{Acknowledgements}
SCA and TW are each supported by an NSERC Discovery research grant.

The referees are thanked for remarks which have improved parts of this work. 

On behalf of all authors, the corresponding author states that there is no conflict of interest. 
Data sharing is not applicable to this article as no datasets were generated or analysed during the current study.

\end{document}